\documentclass[aps,prb,twocolumn,superscriptaddress]{revtex4-1}
\usepackage{graphicx}
\usepackage{picture}
\usepackage{xcolor}
\usepackage{color}
\usepackage{tikz}
\usepackage{ulem}
\usepackage{amsmath}
\usepackage{rotating}

\begin{document}
\title{Nonequilibrium DMFT+CPA for Correlated Disordered Systems}

\author{Eric Dohner}
\affiliation{Department of Physics, University at Albany (SUNY), Albany, New York 12222, USA}
\author{Hanna Terletska}
\affiliation{Department of Physics and Astronomy, Middle Tennessee State University, Murfreesboro, TN 37132, USA}
\author{Ka-Ming Tam}
\affiliation{Department of Physics and Astronomy, Louisiana State University, Baton Rouge, LA 70803, USA}
\affiliation{Center for Computation and Technology, Louisiana State University, Baton Rouge, LA 70803, USA}
\author{Juana Moreno}
\affiliation{Department of Physics and Astronomy, Louisiana State University, Baton Rouge, LA 70803, USA}
\affiliation{Center for Computation and Technology, Louisiana State University, Baton Rouge, LA 70803, USA}
\author{Herbert F Fotso}
\affiliation{Department of Physics, University at Albany (SUNY), Albany, New York 12222, USA}

\begin{abstract}

We present a solution for the nonequilibrium dynamics of an interacting disordered system. The approach adapts the combination of the equilibrium dynamical mean field theory (DMFT) and the equilibrium coherent potential approximation (CPA) methods to the nonequilibrium many-body formalism, using the Kadanoff-Baym-Keldysh complex time contour, for the dynamics of interacting disordered systems away from equilibrium. We use our time domain solution to obtain the equilibrium density of states of the disordered interacting system described by the Anderson-Hubbard model, bypassing the necessity for the cumbersome analytical continuation process. We further apply the nonequilibrium solution to the interaction quench problem for an isolated disordered system. Here, the interaction is abruptly changed from zero (non-interacting system) to another constant (finite) value at which it is subsequently kept. We observe via the time-dependence of the potential, kinetic, and total energies the effect of disorder on the relaxation of the system as a function of final interaction strength. The real-time approach has the potential to shed new light on the fundamental role of disorder in the nonequilibrium dynamics of interacting quantum systems.

\end{abstract}

\maketitle

\section{Introduction}
\label{sec:introduction} 

The physics of strongly correlated systems remains the subject of sustained research efforts due to the many intriguing properties that they exhibit.
Dynamical mean field theory (DMFT) is now well established as an essential tool in advancing the understanding of these systems in equilibrium. \cite{DMFT, DMFT_2, DMFT_3, DMFT_4, DMFT_FK} The method and its cluster extensions \cite{DCA_1, DCA_2, DCA_3, DCA_review1, DCA_review2, CDMFT} have been used extensively for strongly correlated systems. It has been extended to the nonequilibrium problem and used effectively to study the dynamics away from equilibrium in different settings \cite{DMFT_noneq, FK_NonEq_DMFT08, DMFT_noneq_Aoki, thermalization, steadyState1, steadyState2, steadyState_Aoki, NoneqFDT_Frontiers}. Although it has been adapted to the treatment of heterogeneous systems \cite{PotthoffNoltingPRB99, LinOkamotoMillisPRB06, FreericksPRB2004}, the approach is typically focused on clean systems.  However, whether by design or as a result of crystal growth constraints, disorder is ubiquitous and plays a central role in real materials and in the devices that they enable.  
The interplay between disorder and electron-electron interaction strongly influences the electronic structure and transport properties of materials, and is responsible of many unusual phenomena.~\cite{UW1, UW2, UW3} Both disorder and electron interactions are the driving forces for the associated metal-insulator transitions with electron localization resulting from electron-electron interaction~\cite{Mott1,Mott2} or from disorder ~\cite{Anderson1, Anderson2, TMDCA_review}. The presence of both effects gives rise to intriguing behaviors such as many-body localization (MBL) that are the subject of intense activity in relevant research communities.~\cite{MBL1, MBL2, MBL3, MBL4, MBL5}

The coherent potential approximation (CPA), that predates DMFT and shares similarities with this approach in its formulation, has been separately used extensively to study various disordered systems.~\cite{CPA_Soven_1967, CPA_Kirkpatrick, CPA_Velicky_1969, CPA_Yonezawa_1973}
While DMFT maps the lattice problem onto an impurity embedded in a self-consistently determined host, CPA simulates scattering in a random potential by a self-consistently determined homogeneous effective host.
The CPA method has also been extended to the nonequilibrium dynamics of disordered systems and applied to the analysis of transport in various systems. \cite{NonEqCPA_1, NonEqCPA_2} 

Both CPA and DMFT are Green's function-based approaches, and can be easily combined to study the interplay of disorder and electron interactions. \cite{DMFT_CPA_1, DMFT_CPA_2, DMFT_CPA_3, DMFT_CPA_4, DMFT_CPA_5} However, methods that can describe these ever present interplays when systems of interest are driven away from equilibrium are still lacking. 

The implementation of such an approach, combining both CPA and DMFT nonequilibrium solutions, is the focus of the present paper. We implement this solution for interacting disordered systems on the complex time axis. We use our solution to extract the density of states of the equilibrium system for different values of the interaction and different disorder strengths, observing at strong interactions the insulator-to-metal transition that has been reported in certain correlated systems. \cite{insulator_to_metal_1, insulator_to_metal_2} While these results for the density of states are not novel in themselves, obtaining them from our real-time simulation not only enables us to test the validity of our solution, but these calculations of equilibrium density of states from a real time formalism have the added advantage of bypassing the cumbersome analytical continuation process. We further apply the formalism to the interaction quench of an Anderson-Hubbard model where the noninteracting system, initially in equilibrium at finite temperature, has the interaction abruptly switched to another, finite, value at a given time. This process reveals different dynamics in the time-dependent energies across the quench as a function of disorder strength.

The rest of the paper is structured as follows. In section \ref{sec:Model-Methods}, we discuss the model for the interacting disordered system and the nonequilibrium formalism combining both DMFT and CPA. In section \ref{sec:Results}, we present some results for the densities of states of the equilibrium problem and the relaxation of the time-dependent energies for various final interaction strengths as a function of disorder strength. Finally, we end with our conclusion in section \ref{sec:conclusion}.

\section{Model and Methods}
\label{sec:Model-Methods}

\subsection{Model}

We are interested in an interacting disordered system that can be described in equilibrium by the single-band Anderson-Hubbard model defined by:  
    \begin{eqnarray}
    \label{eq:AndersonHubbard}
         H  = - \sum_{\langle i j \rangle \sigma} &t_{ij}& (c^\dagger_{i \sigma} c_{j \sigma} + h.c.) + \sum_{i} U n_{i \uparrow} n_{i \downarrow} \nonumber \\
        & + &  \sum_{i \sigma}\left(V_i - \mu \right) n_{i \sigma},
    \end{eqnarray}
The first term represents the kinetic energy, the second term the interaction $U$ between electrons, $V$ describes the random disorder potential, and $\mu$ is the chemical potential. $t_{ij} = t_{hop}$ is the hopping amplitude between nearest neighboring sites denoted by $\langle i j \rangle$. We work in units where $c = \hbar = e = 1.$
$c^\dagger_{i \sigma}$ ($c_{i \sigma}$) is the creation (annihilation) operator for a particle of spin $\sigma =\uparrow,\downarrow$ at site $i$. $U$ is the Coulomb interaction at a doubly occupied site. $n_{i, \sigma}= c^\dagger_{i \sigma}c_{i \sigma}$ is the number operator for particles of spin $\sigma$ at site $i$. $V_i$ is the local on-site disorder potential randomly distributed according to a probability distribution $P(V_i)$. We use a ``box" distribution $P(V_i)=\frac{1}{2W}\Theta(W-|V_i|)$. We employ the shorthand notation $<...>_{\{V\}}=\int dV_i P(V_i) (...)$ to denote the disorder averaging. Our analysis of the Anderson-Hubbard model will be done on the Bethe lattice with a large coordination number $z  \rightarrow \infty$. 
We will study this system at half-filling when, at time $t_{quench}$, the interaction is abruptly switched from an initial value of $U_1 = 0$ to a final value $U_2 = U$ while the disorder strength remains constant at its set value. 

\subsection{Nonequilibrium Formalism}
For the nonequilibrium many-body formalism, starting at an initial time $t_{min}$, the system is evolved forward in time to times of physical interest up to a maximal time $t_{max}$, and then backwards again to the initial time $t_{min}$.  The formalism involves different types of Green's functions including $G^<(t,t')$  (the lesser),  $G^>(t,t')$ (the greater), and $G^R(t,t')$ (the retarded) Green's functions. Physical observables can be obtained from these different Green's functions.

\begin{figure}[htbp]
\includegraphics*[width=6.0cm, height=5.0cm]{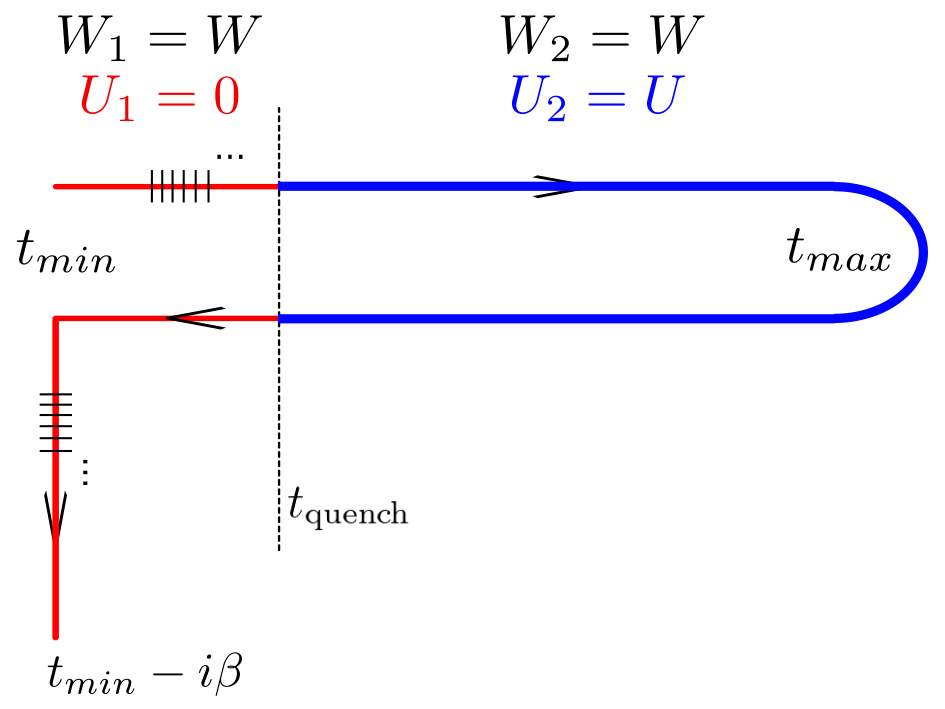}
\caption{The Kadanoff-Baym-Keldysh contour, with initial and final times $t_{min}$ and $t_{max}$. The interaction quench with the interaction being switched from $U_1=0$ to a finite value $U_2=U$ occurs at time $t_{quench}$. The disorder strength $W$ is held fixed. The real-time and imaginary-time parts of the contour are discretized with respective step sizes $\Delta t$ and $\Delta \tau$.}
\label{fig:keldyshcontour}
\end{figure}

For a system initially in equilibrium at a temperature $T=1/\beta$, a vertical branch of imaginary times is added to the time evolution resulting in the Kadanoff-Baym-Keldysh contour\cite{Keldysh64_65, BaymKadanoff62, StefanucciLeeuwen_CUP2013, rammer_2007}. This adds to the different Green's functions, the Matsubara Green's function and the mixed time Green's functions, for which one of the two times is on either the forward or the backwards horizontal branch of real times, while the other is on the vertical branch of imaginary times. The time evolution on the contour is illustrated schematically in FIG.\ref{fig:keldyshcontour}. In general, the formalism can be either formulated explicitly in terms of the different Green's functions or in terms of the contour-ordered Green's function from which all others can be extracted. In the latter situation, which we adopt in the present paper, the formalism has the advantage of being very similar to that of the equilibrium problem.\\
The contour-ordered Green's function is given by: 
\begin{equation}
    G^c_{i, j, \sigma}(t, t')  =  \theta_c(t, t')G^>_{i,j, \sigma}(t, t') + \theta_c(t', t)G^<_{i,j, \sigma}(t, t'). \;\;\;\;\label{eq:GContour} 
\end{equation}
With the lesser and greater  Green's functions defined by operator averages in the Heisenberg representation:
\begin{eqnarray}
 G^<_{i,j, \sigma}(t, t') & = &  i\langle c^{\dagger}_{i \sigma}(t') c_{j \sigma}^{\phantom\dagger}(t)\rangle, \label{eq:GLesser}  \\
 G^>_{i,j, \sigma}(t, t') & = & -i\langle c_{i \sigma}^{\phantom\dagger}(t) c^{\dagger}_{j \sigma}(t')\rangle. \label{eq:GGreater}  
\end{eqnarray}
 From these, we can construct the retarded and advanced Green's functions that are defined by:
 \begin{eqnarray}
 G^R_{i,j, \sigma}(t,t') & = & -i\theta(t-t')\langle \{c_{i\sigma}^{\phantom\dagger}(t), c^{\dagger}_{j\sigma}(t')\}\rangle \label{eq:GRetarded}. \\
 G^A_{i,j, \sigma}(t, t') & = & i\theta(t'-t)\langle \{c_{i\sigma}^{\phantom\dagger}(t), c^{\dagger}_{j\sigma}(t')\}\rangle.\label{eq:GAdvanced}
\end{eqnarray}
$\theta_c(t,t')$ is the contour-ordered Heaviside function. It orders time with respect to the contour: it is equal to $1$ if $t$ is ahead of $t'$ on the contour and is equal to $0$ otherwise. Hereafter we drop the superscript $c$ on the contour-ordered Green's function: any correlation function not identified as a particular type (e.g., not $G^<$) should be understood to refer to the full contour-ordered Green's function.

In these expressions, $c^{\dagger}_{i \sigma}(t)$ and $c_{i \sigma}(t)$ are, respectively, the Heisenberg representation of the creation and the annihilation operators for an electron at site $i$ with spin $\sigma$ at time $t$; $\theta$ is the usual Heaviside function ({\it i.~e.},~$\theta(t-t') = 0$ if $t<t'$ and $\theta(t, t') = 1$ otherwise); $\{\mathcal{A}, \mathcal{B}\}$ is the anticommutator of operators $\mathcal{A}$ and $\mathcal{B}$. The symbol $\langle \mathcal{A} \rangle$ is the expectation value of the operator $\mathcal{A}$ evaluated with respect to the initial thermal state:
\begin{equation}
    \langle \mathcal{A} \rangle=\frac{{\rm Tr} e^{-\beta\mathcal{H}(t_{\rm min})}\mathcal{A}}{{\rm Tr} e^{-\beta\mathcal{H}(t_{\rm min})}},
\end{equation}
where $\mathcal{H}(t_{\rm min}) = {\mathcal H}_{eq}$ is the initial equilibrium Hamiltonian before the quench.

We will particularly examine the dynamics of the system through the time evolution of the total, potential and kinetic energy of the system when the interaction is abruptly quenched from $U_1 =0$ to a finite value $U_2 = U$.

Within the DMFT framework, starting from the lattice action and integrating out all sites except site $i$, the effective action for the considered Hamiltonian can be written as:
\begin{eqnarray}
 \label{eq:AH_effectiveAction}
 S_{eff} &=& -i\sum_{\sigma}\int_{\mathcal{C}} dt dt' c^{\dagger}_{\sigma, i}(t) \Delta(t,t') c_{\sigma, i}(t') \nonumber \\
 &-&i\int_{\mathcal{C}} dt H_{loc}(t)
\end{eqnarray}
Where $H_{loc}$ is the local part of the Hamiltonian at site $i$ that includes the disorder value at this site. In an extension of the equilibrium CPA, the hybridization $\Delta(t,t')$ is obtained from the disorder averaged local Green's function. From this action, we are readily able to extend the equilibrium treatment of disorder and interaction \cite{DMFT_CPA_1, DMFT_CPA_2, DMFT_CPA_3, DMFT_CPA_4} to the nonequilibrium problem.

The impurity Green's function for a given disorder configuration is given by:
\begin{eqnarray}
 \label{eq:ImpurityG}
 G_{V_i}(t,t') & = & -i\langle c(t) c^{\dagger}(t') \rangle_{S_{eff}} \\
         & = & \left[ \mathcal{G}_{V_i}^{-1}(t,t') - \Sigma_{V_i}(t,t') \right]^{-1}.
\label{eq:ImpurityG2}
\end{eqnarray}
Where $\Sigma_{V_i}(t,t')$ is the interaction self-energy for the disorder configuration of the action (\ref{eq:AH_effectiveAction}) and captures effects of the interaction on the impurity. $\mathcal{G}$ is non-interacting Green's function for the impurity problem given by:
\begin{equation}
 \label{eq:Gscript}
 \mathcal{G}_{V_i}(t,t') = \left( \left(i\partial_t + \mu -V_i \right)\delta(t,t') - \Delta(t,t') \right)^{-1}
\end{equation}
In this way, the disorder averaged local Green's function $G_{ave}$ is obtained by averaging the impurity Green's function over all disorder configurations:
\begin{equation}
 \label{eq:GlocalAverage}
 G_{ave}(t,t') = \langle \left( \mathcal{G}_{V_i}^{-1}(t,t') - \Sigma_{V_i}(t,t') \right)^{-1} \rangle_{\{V\}}
\end{equation}
The symbol $\langle \cdot \cdot \cdot \rangle_{\{V\}}$ denotes average over all possible disorder configurations, which we perform by taking the numeric integral of $\int dV P(V) \cdot \cdot \cdot $ using the midpoint rectangular rule.

On the infinite dimensional Bethe lattice, the hybridization is then expressed as:
\begin{equation}
 \label{eq:hybridization}
 \Delta(t,t') = t^{*^2} G_{ave}(t,t').
\end{equation}
Where $t^*$ is the hopping amplitude rescaled with the coordination number $z$
so that $ t_{hop} = \frac{t^*}{\sqrt{z}} $. We use $t^* = 0.25$ and thus set the bandwidth to be our energy unit and its inverse to be the time unit. 
In our solution of (\ref{eq:ImpurityG2}), we use second order perturbation theory as the impurity solver, giving us the self-energy:
\begin{equation}
\label{eq:sigmaU} 
 \Sigma_{V_i}(t,t') = -U(t)U(t') \mathcal{G}_{V_i}(t,t')^2 \mathcal{G}_{V_i}(t',t).
\end{equation}
This self-energy $\Sigma_{V_i}$ is the interaction self-energy and it is calculated, at this stage, for a specific disorder configuration. Given that the disorder is symmetric, the chemical potential is set to $\mu=U/2$ and dropping the Hartree term ensures half-filling $(n=1)$ since we use the disorder averaged Green's function to set the filling.


\begin{figure}[htbp]
\includegraphics*[scale = 0.55]{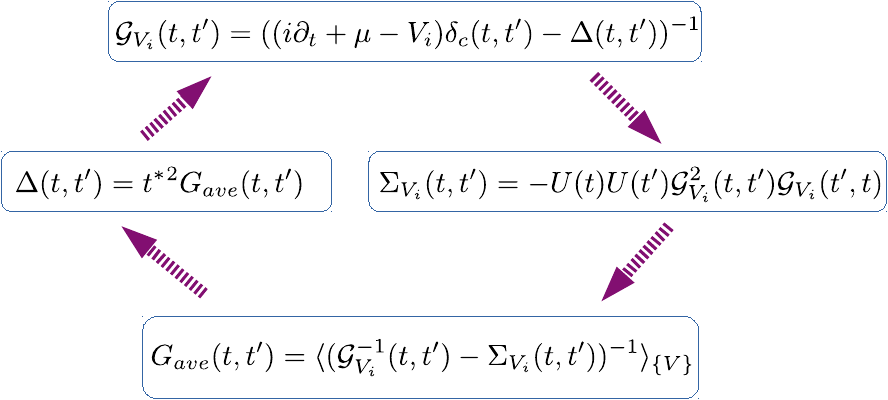}
\caption{Self-consistency loop for the nonequilibrium DMFT+CPA algorithm on the Bethe lattice.} 
\label{fig:selfConsistencyLoop}
\end{figure}

It should be noted that this nonequilibrium formalism clearly reduces to DMFT for the clean/non-disordered system $(W=0)$ and to the equilibrium CPA for a non-interacting system $(U=0)$ in equilibrium.

\subsection{Algorithm and Numerical Implementation}

The nonequilibrium DMFT+CPA algorithm follows the self-consistency loop illustrated in FIG. (\ref{fig:selfConsistencyLoop}). The loop is started by setting the hybridization $\Delta(t,t')$ to an initial guess (we use an infinitesimal imaginary number) for the first calculation of the non-interacting Green's function on the impurity (\ref{eq:Gscript}). From this, the self-energy $\Sigma_{V_i}$ of (\ref{eq:sigmaU}) is calculated for each configuration of the disorder, and the average Green's function $G_{ave}$ of (\ref{eq:GlocalAverage}) is calculated by averaging over all disorder configurations. At each subsequent iteration, the new hybridization is calculated from the average Green's function by (\ref{eq:hybridization}). This process is repeated until convergence of the average Green's function within a desired criteria.

Our implementation of the contour-ordered Green's function follows that of Ref.[\onlinecite{FK_NonEq_DMFT08}]. The different matrices are represented as square complex matrices of size $(2N_t + N_{\tau}) \times (2N_t + N_{\tau})$ with each index representing a point along the complex time axis. $N_t$ is the number of points along each real-time branch of the contour; $N_{\tau}$ is the number of points along the imaginary-time branch. Certain observables (the energies in particular) need to be extrapolated to the $\Delta t \rightarrow 0$ limit. To this end, the calculation is performed for multiple values of $\Delta t$, and standard Lagrange interpolating polynomials are applied to obtain the $\Delta t \rightarrow 0$ values of observables. In this paper, we use $N_{\tau} = 200$ and typically, $N_t$ values of $800, 1000$ and $1200$ or $1000, 1200$ and $1400$ followed by an extrapolation to the $\Delta t \rightarrow 0 $ limit using Lagrange polynomials on the 3 time grids.

We use the rectangular leftpoint integration rule, such that 
\begin{equation}
    \label{eqn:leftpointdiscretization}
    \int_C dt F(t) \rightarrow \sum_{i = 1}^{2N_t + N_\tau} w_i F_i.
\end{equation}
Where $w_i$ is the integral weight, defined by:
\begin{equation}
\begin{aligned}
    \label{eqn:integralweights}
    w_i &= ~\Delta t,~~~ 1 \leq i \leq N_t \\
        &= -\Delta t,~~ N_t < i \leq 2N_t \\
        &= - i \Delta \tau,~ 2N_t < i < 2N_t + N_\tau.
\end{aligned}
\end{equation}
The delta function on the contour can be discretized in multiple equivalent ways. Following Ref.[\onlinecite{FK_NonEq_DMFT08}], it is often convenient to point-split the delta function, such that the first subdiagonal is occupied rather than the diagonal:
    \begin{equation}
        \label{eqn: deltaij}
        \delta(t_i, t_j) \rightarrow \frac{\delta_{i j+1}}{w_i}
    \end{equation}
The product of two Green's functions in frequency space becomes a convolution in contour time, which when discretized is evaluated as a matrix product weighted by the $w_i$'s:
    \begin{equation}\label{eqn:numericsproduct}
        [A*B](t,t') = \int_C d\bar{t} A(t,\bar{t}) B(\bar{t}, t') \rightarrow \sum_{k = 1}^{2N_t + N_\tau} A_{ik} w_k B_{kj}
    \end{equation}
and the continuous matrix inverse becomes a discrete matrix inverse. Appropriately including the definition of the delta function yields
    \begin{equation}\label{eqn:numericsinverse}
        A^{-1} \rightarrow [w_i A_{ij} w_j]^{-1}.
    \end{equation}
Where the quantity in square brackets is a discrete matrix. Both Eqns.(\ref{eqn:numericsproduct}) and (\ref{eqn:numericsinverse}) can then be evaluated by standard linear algebra routines such as LAPACK.

Equations (\ref{eq:GlocalAverage}) and (\ref{eq:sigmaU}) can be efficiently parallelized with the number of parallel processes defined by the number of points on the integration over possible disorder configurations. In practice, a few hundred points at most are sufficient for the disorder type that we consider.

\begin{figure}[t] 
\includegraphics[width= 6cm, height=6cm]{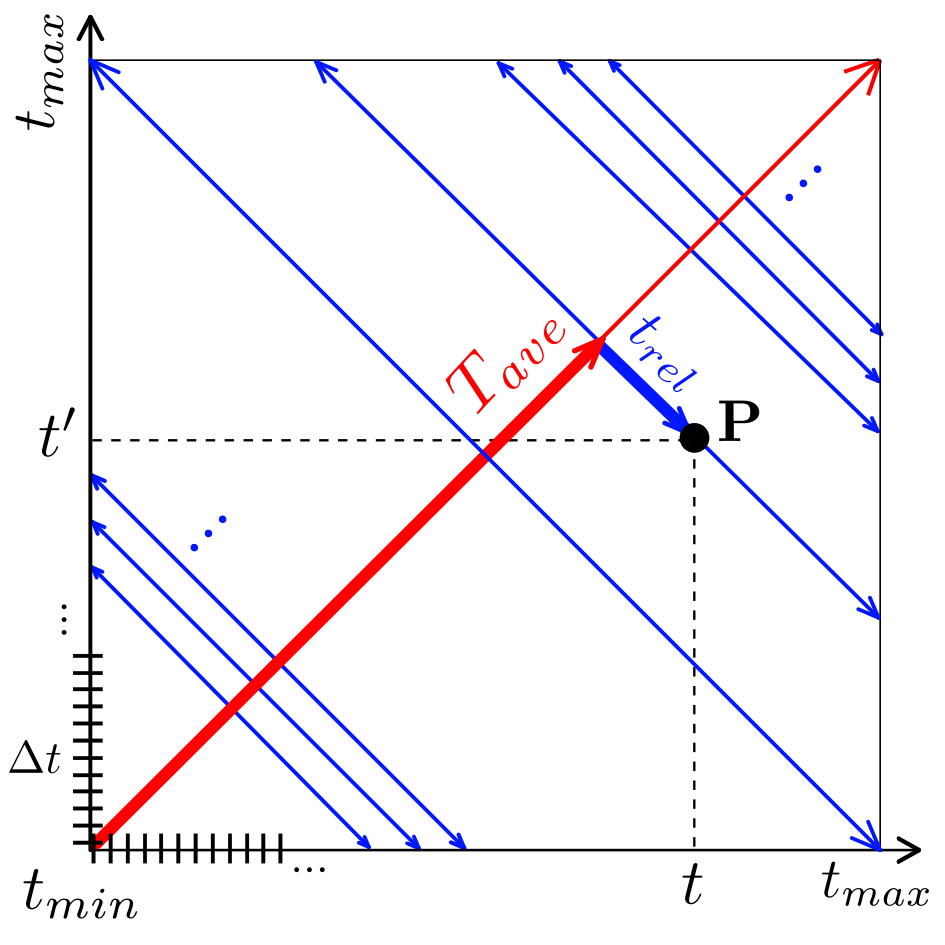}
\caption{Illustration of the relation between the contour time coordinates $(t,t')$  and the Wigner time coordinates $(T_{ave}, t_{rel})$ of the point $P$ in the \textit{two-time} space.}
\label{fig:wignerCoordinates}
\end{figure}

\begin{figure}[t] 
\begin{center}
 \includegraphics[width=8cm, height=6cm]{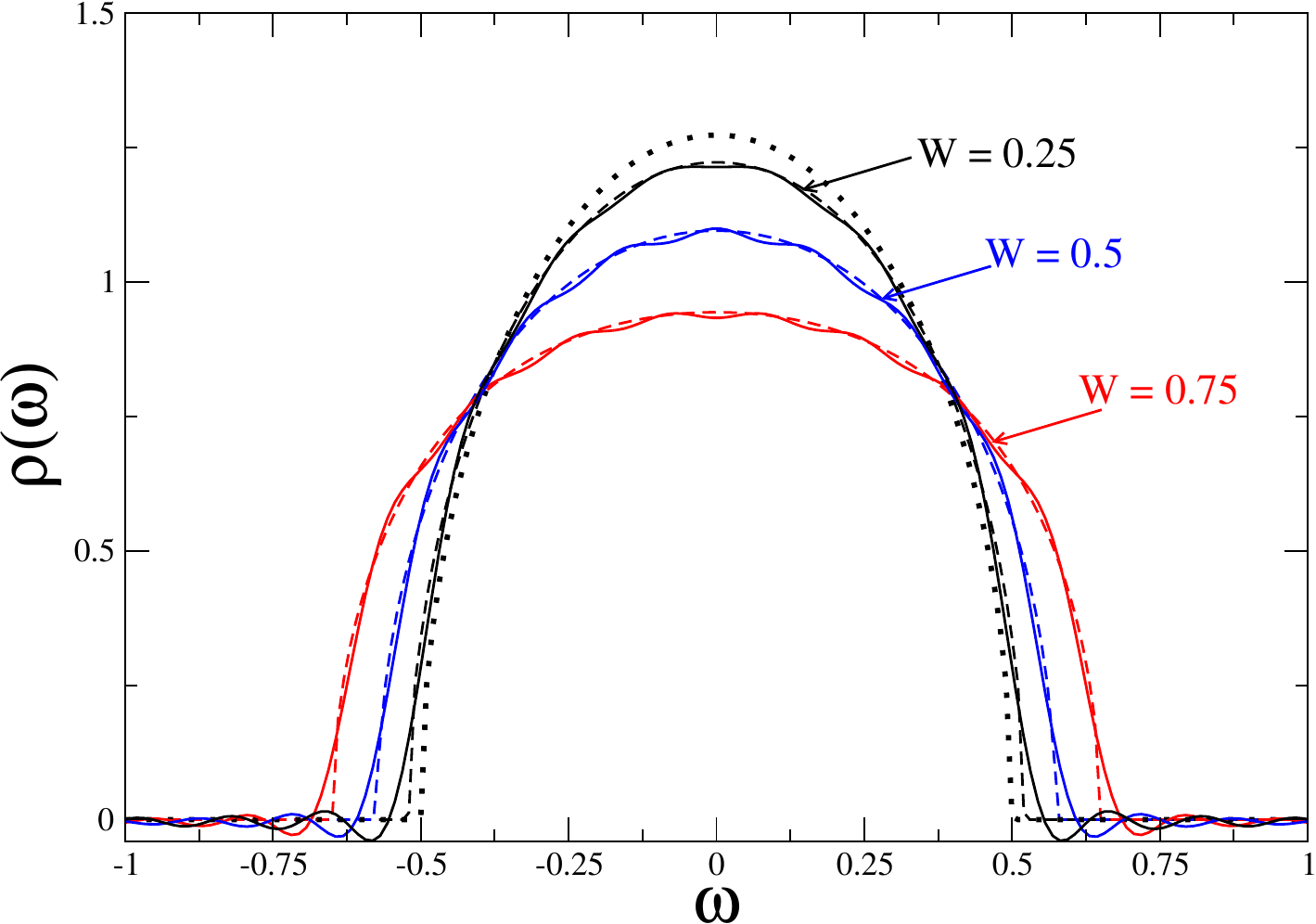}
  \caption{Equilibrium densities of states for the non-interacting $(U=0)$ tight-binding Anderson model at three values of disorder strength $W=0.25, 0.5, 0.75$. At a given disorder strength $W$, dashed lines are obtained using the standard frequency-space CPA calculation formalism, while the solid lines are obtained via the Fourier transform of the Green's function calculated using the contour-time formalism. The dotted black line is the density of states for non-interacting clean system ($W=0, U=0$). Oscillations in the solid lines are due to the Gibbs phenomenon in the Fourier transform of the time domain solution.}
\label{fig:andersonDOSwt}
\end{center}
\end{figure}

\section{Results}
\label{sec:Results}
\subsection{Equilibrium density of states}

To demonstrate the validity of the developed non-equilibrium DMFT+CPA method, we first apply the nonequilibrium formalism described above for the equilibrium system in real time. We compare the results from our time-dependent approach with those calculated using a real-frequency equilibrium approach.
In this context, the density of states can be obtained from the nonequilibrium retarded Green's function $G^R(t,t')$. First, the time coordinates are changed from $(t, t')$ to the Wigner coordinates $(T_{ave}, t_{rel})$. The $(t,t') \to (T_{ave}, t_{rel})$ change of coordinates is schematically illustrated in FIG.\ref{fig:wignerCoordinates}. $T_{ave}$ is typically viewed as the effective time of the system while $t_{rel}$ is the time with respect to which Fourier transforms are performed to obtain frequency space quantities. A Fourier transform on this form of the retarded Green's function: $G(T_{ave}, \omega) = \int dt_{rel} ~e^{i \omega t_{rel}}~ G(T_{ave},t_{rel})$ yields the density of states $\rho(\omega) =-\frac{1}{\pi} \mathrm{Im}\left[ G^R(T_{ave}, \omega)\right]$.
For the equilibrium system, the result of this operation is, in principle, independent of $T_{ave}$. However, a choice has to be made for a value of $T_{ave}$ at which the range of $t_{rel}$ values available enables the best numerical evaluation of the Fourier transform for the density of states (typically halfway along the average time axis). 

FIG.\ref{fig:andersonDOSwt} shows the density of states $\rho(\omega)$ obtained from the time domain nonequilibrium approach for the  Anderson model of non-interacting
electrons subjected to a random disorder potential (Hamiltonian of Eq.~(\ref{eq:AndersonHubbard}) with $U=0$). The solution here corresponds to the nonequilibrium CPA. For the clean noninteracting system ($W=0$ and $U=0$), the density of states has a semi-elliptical lineshape $\rho_{0}(\omega) = \frac{1}{2\pi t^{*2}} \sqrt{4t^{*2}-\omega^2}$
(dotted line in FIG.\ref{fig:andersonDOSwt}). For a given disorder strength $W$, the density of states obtained from the time domain calculation (solid line) is compared with the frequency domain (dash line) CPA results. As expected, increasing the disorder strength $W$ causes the broadening and suppression of the spectral peak. The density of states for this noninteracting system has sharp edges that are hard to resolve numerically and give rise to oscillations due to Gibbs phenomenon in the Fourier transform of the real-time approach. Nevertheless, the overall lineshape is in good agreement between the two methods.

\begin{figure}[t] 
\begin{center}
\includegraphics*[width=8.0cm]{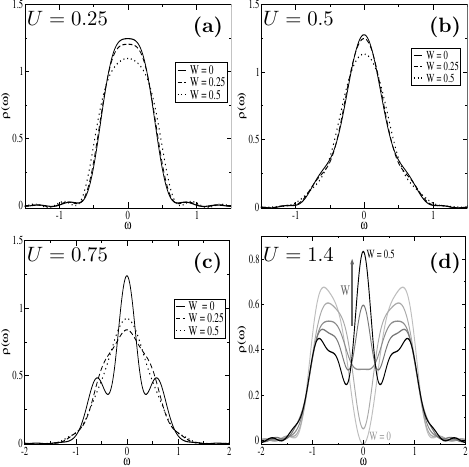}
    \caption{Equilibrium densities of states with different disorder strengths for the equilibrium Anderson-Hubbard model at temperature $T$ such that $1/T=\beta=40$ with $U=0.25$ (a), $U=0.5$ (b), $U=0.75$ (c), $U=1.4$ (d). In (a), (b) and (c), the solid line represents the clean system limit $W = 0$, the dashed line $W=0.25$, and the dotted line $W=0.5$.  In panel (d), a gray scale is used for disorder strengths varying from $0$ (lightest shade) to $0.5$ (darkest shade).} 
\label{fig:2x2DOS}
\end{center}
\end{figure}

We now apply the method to the Anderson-Hubbard model in equilibrium at temperature $T$ such that $1/T=\beta=40$ for different parameters and examine the densities of states obtained from our real-time method. These results are presented in FIG.\ref{fig:2x2DOS}  for the equilibrium interacting disordered model with $U=0.25$ (a), $U=0.5$ (b), $U=0.75$ (c), $U=1.4$ (d). In panels (a), (b) and (c), the solid line represents the clean system limit $W = 0$, the dashed line $W=0.25$ and the dotted line $W=0.5$. In panel (d), a gray scale is used with darker shades indicating stronger disorder. The clean system restores the expected equilibrium Hubbard density of states for the Bethe lattice.\cite{Bulla99} For weak interactions (FIG.\ref{fig:2x2DOS}-(a,b)), as the disorder is tuned from weak to strong values, we observe a broadening of the density of states. For moderate interaction strength (FIG.\ref{fig:2x2DOS}-(c)), the density of states at weak disorder display Hubbard sidebands and a quasiparticle peak. As the disorder strength is increased, the sidebands and the quasiparticle peak are suppressed in favor of a single broad peak akin to the density of states of the weakly interacting clean system. This behavior is more pronounced at strong interactions (FIG.\ref{fig:2x2DOS}-(d)) where the the clean system displays a gap separated by the two Hubbard bands. Increasing the disorder strength for this system gradually fills the gap in a process similar to the insulator-to-metal transition that has been reported in certain correlated materials. \cite{insulator_to_metal_1, insulator_to_metal_2}

\begin{figure}[t] 
\begin{center}
\includegraphics*[width=8.0cm]{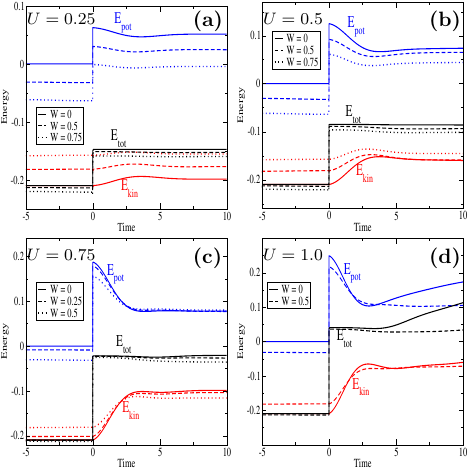}
    \caption{Relaxation, as a function of disorder strengths, of the kinetic, potential, and total energies of the Anderson-Hubbard model when the interaction is switched at time $t=0$ from $U_1=0$ to a finite value of $U_2=U$. In panels (a) for $U=0.25$ and (b) for $U=0.5$, the relaxation is shown for the clean system $W = 0$ (solid line), $W = 0.5$ (dashed line), and $W=0.75$ (dotted line). In panel (c) for $U=0.75$, results are shown for the clean system $W = 0$ (solid line), $W = 0.25$ (dashed line), and $W=0.5$ (dotted line). In panel (d) for $U=1.0$, results are shown for the clean system $W = 0$ (solid line), $W = 0.5$ (dashed line). Panel (d) for $U=1.0$ shows the breakdown of the solution with second order perturbation theory as the impurity solver.}
\label{fig:2x2energies}
\end{center}
\end{figure}

\subsection{Relaxation of the energy across the interaction quench}

We study the time-evolution of the kinetic, potential and total energy in time when the interaction is suddenly switched on from an initial non-interacting system ($U_1=0$) in thermal equilibrium at temperature $T$ given by $1/T = \beta = 40$ to a finite interaction strength $U_2=U$.

The total energy is obtained by summing up the kinetic and potential energies\cite{HubbardQuenchEckstein}.

The kinetic energy per lattice site is defined by:
\begin{equation}
E_{kin}(t) = \frac{1}{N} \sum_{k, {\sigma}} \epsilon_k \langle c^{\dagger}_{k, \sigma}(t) c_{k, \sigma}(t) \rangle. 
\end{equation}
Where $N$ is the number of sites, $k$ is the momentum vector, and $\epsilon_k$ is the dispersion relation. The kinetic energy can thus be rewritten as:
\begin{equation}
E_{kin}(t) = 2\int \rho(\epsilon) \epsilon G^<_{\epsilon}(t,t).
\end{equation}
Where $\epsilon$ is the band energy.\\
The potential energy follows from the expression of the double occupancy:
\begin{equation}
E_{pot}(t) = [G_{ave}*\Sigma_{ave}]^<(t,t) + \frac{U(t)}{4}.  
\label{eq:Epot}
\end{equation}


Where
\begin{equation}
G^<_\epsilon(t,t') = \left\{ [(i\partial_t - \epsilon) \delta(t,t') - \Sigma(t,t')]^{-1} \right\}^<
\label{eq:GlesserLattice}
\end{equation}
is the lattice lesser Green's function, $G_{ave}$ is the Green's function averaged over all disorder configurations, $\Sigma_{ave}$ is the self-energy obtained from the Dyson equation with $G_{ave}$, and the noninteracting Green's function. It thus includes the effects of both the interaction and the disorder. The lesser part of the convolution  $G_{ave}*\Sigma_{ave}$ is taken in equation (\ref{eq:Epot}) for the potential energy.


Since the system is isolated, the total energy has a constant value before the quench and another constant value after. For the equilibrium system before the interaction quench, both the potential and kinetic energy are also constant. However, they exhibit nontrivial dynamics after the quench. 
This is illustrated in FIG.\ref{fig:2x2energies}. In panels (a) for $U=0.25$ and (b) for $U=0.5$, the relaxation is shown for the clean system $W = 0$ (solid line), $W = 0.5$ (dashed line), and $W=0.75$ (dotted line). In panel (c) for $U=0.75$, results are shown for the clean system $W = 0$ (solid line), $W = 0.25$ (dashed line), and $W=0.5$ (dotted line). In panel (d) for $U=1.0$, results are shown for the clean system $W = 0$ (solid line), $W = 0.5$ (dashed line). 
For weak final interactions where the density of states is broadened by the disorder, the relaxation of both the potential and kinetic energy have a monotonic evolution before a plateau at their steady state values. This is shown in FIG.\ref{fig:2x2energies}-(a) for $U=0.25$ and FIG.\ref{fig:2x2energies}-(b) for $U=0.5$. Here, the steady state kinetic energy increases with disorder strength while the steady state potential energy decreases with increasing disorder strength. FIG.\ref{fig:2x2energies}-(c) shows the energy relaxation as the final interaction is increased to moderately strong values ($U=0.75$) where the equilibrium density of states would feature a quasiparticle peak flanked by the onset of the Hubbard sidebands. The steady state potential energy shows little change with the disorder strength while the steady state kinetic energy decreases with increasing disorder strength.

Our solution for the nonequilibrium problem using second order perturbation theory as an impurity solver breaks down for strong interactions for this interacting disordered system. This breakdown is manifested through the divergence of the kinetic and potential energies and a total energy after the quench that is not constant as pictured in FIG.\ref{fig:2x2energies}-(d) for $U=1.0$. The breakdown is similar to what was previously observed for the interaction quench of the clean system using nonequilibrium DMFT with second order perturbation theory as an impurity solver.\cite{HubbardQuenchEckstein} Note also that the self-consistency loop for both the equilibrium and the nonequilibrium situation becomes unstable for strong interaction strengths and for strong disorder ($U$ and $W$ of the order of the bandwidth). Nevertheless, our solutions are robust for weak to moderate interaction strengths. \\

\begin{figure}[t] 
\begin{center}
\includegraphics*[width=8.0cm]{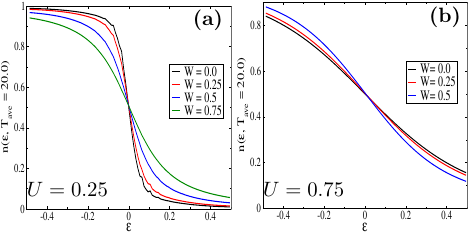}
    \caption{Momentum distribution function at the latest average time of the simulation for the initial non-interacting equilibrium system at a temperature such that $1/T=\beta=40$ quenched at time $t=0$ to an interaction strength $U=0.25$ (a), $U=0.75$ (b) for disorder strengths $W=0.0$ (black), $W=0.25$ (red), $W=0.5$ (blue) and $W=0.75$ (green line in panel (a) only).} 
\label{fig:2x1distrFunct}
\end{center}
\end{figure}

We further analyse the relaxation of the system after the quench by plotting the momentum distribution function as a function of disorder strength for the two observed relaxation scenarios at a late simulation time, $n(\epsilon, t_{ave}=20.0) = G^<_{\epsilon}(t,t)$. Where $G^<_{\epsilon}(t,t)$ is the equal time lesser Green's function from equation (\ref{eq:GlesserLattice}). This momentum distribution is shown in FIG.\ref{fig:2x1distrFunct} for $U=0.25$ (a) and $U=0.75$ (b).  Our results show that the momentum distribution function for an interaction quench on a system that is initially non-interacting into weakly interacting system, behaves as if increasing disorder strength is analogous to lowering the temperature FIG.\ref{fig:2x1distrFunct} (a). On the other hand, when the quench takes the system from non-interacting to a moderate interaction strength the momentum distribution after the quench behaves as if the temperature of the system increases with increasing disorder strength FIG.\ref{fig:2x1distrFunct} (b). This is consistent with the identification of the insulator-to-metal transition and of the nontrivial relaxation across the quench as a function of disorder strength when the interaction strength is increased from weak to moderate. Note the absence of a $W=0.75$ curve in FIG.\ref{fig:2x1distrFunct}-(b) because of the breakdown of the formalism discussed above.

\section{Conclusion}
\label{sec:conclusion}
We have presented a nonequilibrium solution for correlated disordered systems using a combination of CPA and DMFT on the complex time axis of the Kadanoff-Baym-Keldysh contour. The solution maps the lattice onto an impurity embedded in a self-consistently determined mean-field and the disorder is treated through averaging over different individual configurations. We applied the approach to the equilibrium problem and showed that it effectively produces the densities of states of the system, bypassing the need for the analytical continuation calculation. To demonstrate the application of the formalism on a nonequilibrium problem, we simulate an interaction quench on the Anderson-Hubbard model. Here, a system initially in equilibrium at finite temperature sees its interaction strength abruptly changed from zero to another finite value. We identify different relaxation processes in the energies of the system as a function of disorder strength. Our solution uses second order perturbation theory as an impurity solver and breaks down at stronger interaction values. We plan to extend these studies to other parameter regimes by adopting other diagrammatic solutions for the impurity solver and also analyze in greater detail the nature of these relaxation processes. Altogether, the approach presents a valuable tool for studies of nonequilibrium dynamics of correlated disordered systems and may shed new light on non trivial dynamics that arise from combined effects of both correlations and disorder when these systems are driven away from equilibrium.

\section*{Acknowledgments} 
HFF is supported by the National Science Foundation under Grant No. PHY-2014023. 
 HT has been supported by NSF OAC-1931367 and NSF
 DMR-1944974 grants. KMT is supported by NSF DMR-1728457 and NSF OAC-1931445. JM is partially supported by the U.S. Department of Energy, Office of Science, Office of Basic Energy Sciences under Award Number DE-SC0017861.

\end{document}